\documentclass[fleqn,usenatbib]{mnras}

\usepackage[T1]{fontenc}

\DeclareRobustCommand{\VAN}[3]{#2}
\let\VANthebibliography\thebibliography
\def\thebibliography{\DeclareRobustCommand{\VAN}[3]{##3}\VANthebibliography}

\usepackage{graphicx}	
\usepackage{amsmath}	
\usepackage{amssymb}	
\usepackage{caption}
\usepackage{subfigure}
\usepackage{booktabs}
\usepackage{longtable}
\usepackage{tabularx}
\usepackage{graphicx}
\usepackage{multirow}

\title[]{Evolutionary tracks of massive stars with different rotation and metallicity in neutrino H-R diagram}

\author[Wang et al.]{Hao Wang$^{1,2}$, Chunhua Zhu$^{1}$\thanks{E-mail:
chunhuazhu@sina.cn}, Helei Liu$^{1}$, Sufen Guo$^{1}$, Guoliang L\"{u}$^{2,1}$\thanks{E-mail: guolianglv@xao.ac.cn}\\
$^{1}$School of Physical Science and Technology, Xinjiang University, Urumqi, 830017, China\\
$^{2}$Xinjiang Astronomical Observatory, Chinese Academy of Sciences, 150 Science 1-Street, Urumqi, 830011, China\\
}

\date{Accepted XXX. Received YYY; in original form ZZZ}

\pubyear{2023}

\begin{document}
\label{firstpage}
\pagerange{\pageref{firstpage}--\pageref{lastpage}}
\maketitle

\begin{abstract}
Neutrino losses play a crucial role in the evolution of massive stars. We study the neutrino luminosity of stars ranging from 20 to 90 $\rm M_{\odot}$ from Zero Age Main Sequence (ZAMS) to Fe Core Collapse (FeCC) with different rotation and metallicity in a neutrino Hertzsprung-Russell diagram. In our simulations, we consider $\rm \omega/\omega_{crit}= 0$ and 0.7 to represent non-rotation and high rotation, respectively, and set the metallicities to 0.014, 0.001, and 0.0001. 
During hydrogen burning stages, neutrino luminosity primarily originates from CNO cycle, and increases with higher stellar mass while decreasing with increasing metallicity. 
For the high metallicity models ($Z=0.014$) during the helium burning stage, the reduction of the hydrogen envelope caused by a larger mass loss rate leads to a gradual decrease in neutrino luminosity.
The rapid rotation results in extra mixing inside massive stars, which increases the neutrino luminosity during main sequence (MS), while decreases the neutrino luminosity during helium burning phase. Simultaneously, the rapid rotation also increases CO core mass, which enhances the neutrino luminosity during C and O burning phase.   
We also investigate the effect of neutrino magnetic moment (NMM) on the massive stars. We find that the energy loss caused by the NMM does not have effects on the evolutionary destiny of massive stars, and it does not significant change the compactness at the time of Fe core collapse. 
\end{abstract}


\section{Introduction}
Massive stars, defined as those with a mass greater than 8 M$_\odot$, undergo a series of nuclear fusion reactions. These include hydrogen burning via the CNO cycle \citep{Bahcall1989}, helium burning via the triple-alpha process, as well as carbon, oxygen, and silicon burning via the alpha process. These reactions lead to the production of heavier elements and eventually to iron core collapse, which can result in a supernova explosion \citep{Woosley2002}. During hydrogen and helium burning, energy is radiated away via photons \citep{Kippenhahn2013}, while during carbon, oxygen, and silicon burning, neutrino cooling becomes the dominant energy loss mechanism due to the rapid increase in internal temperature and density \citep{Raffelt2012}.
Gaining a profound understanding of the critical role of neutrino losses in the evolutionary path of massive stars is indispensable for the progression of our comprehension of astrophysics \citep{Sukhbold2016}.

Neutrino losses are sensitive to temperature and therefore play an important role in the late evolution of massive stars \citep{Kutschera2009, Kato2020a, Fuller2022}. The neutrino delayed heating mechanism is also an important factor leading to supernova explosions \citep{Couch2013, Mirizzi2016, Janka2017}. Due to their extremely small scattering cross section, neutrinos can penetrate freely from the interior of the star and provide valuable insights into the temperature, density, and fusion processes in the core \citep{Farmer2016, Raj2020}. Pre-supernova neutrinos can even serve as an early warning of an impending supernova explosion \citep{Nagakura2022}. Recent studies have also shown that neutrino signals can be used to estimate the radius and mass of neutron stars \citep{Nagakura2022}.

Neutrino detection is rapidly developing as a new astronomical messenger, with advanced neutrino detectors such as Super Kamiokande (SK) \citep{Ikeda2007}, Borexino \citep{Bellini2014}, KamLAND \citep{Eguchi2003}, Irvine-Michigan-Brookhaven(IMB) \citep{Bionta1987} designed to detect the burst of neutrinos emitted during supernova explosions. Borexino \citep{Bellini2014}, SNO+ \citep{Albanese2021}, and SK-GD \citep{MartiMagro2022} play an important role in solar neutrino observations.
While many new generation neutrino detectors, such as JNE \citep{Beacom2017}, DUNE \citep{Abi2020}, JUNO \citep{Abusleme2021}, aim to explore new neutrino events.

Usually, neutrino astronomy mainly focuses on the solar neutrino,
supernova neutrino, even or blazar neutrino. However, it is well known that stars radiate their energy not only by releasing photons from the stellar
surface, but also by radiating neutrinos from the stellar interior.
Very recently, \cite{Shi2020} and \cite{Farag2020} calculated the neutrino luminosity for stars during
different evolutional stages. The former calculated the neutrino luminosity of stars with initial masses ranging from 1 to 9 $\rm M_{\odot}$ during the main sequence (MS) to asymptotic giant branch (AGB) stages. The latter gave the evolutional tracks in a neutrino Hertzsprung-Russell (H-R) diagram for stars with initial mass lower than 40 M$_\odot$.
 
In fact, the stellar structure and evolution depends on stellar rotation and metallicity.
Rotation can have a significant impact on the evolutionary tracks of massive stars, by lowering the effective gravity $g_{\mathrm{eff}}$ and enhancing the luminosity, increases the mass loss rates \citep{Maeder2000}. More importantly, rotational mixing can change the size of stellar core, which may influence neutrino luminosity. At lower metallicity, lower mass and angular momentum loss can make rotational mixing more efficient so that massive star with high rotation ($\omega/\omega_{\mathrm{crit}}=0.7$) can experience quasi-chemically homogeneous evolution \citep{Brott2011,Georgy2013}, it can cause a larger helium core mass and CO core mass \citep{limongi2018}. Therefore it is necessary to include rotation and metallicity when taking about neutrino luminosity.

Meanwhile, many studies have shown that neutrino could have magnetic moment because of its non-zero mass \citep{Marciano1977,Lee1977,Fujikawa1980,Bell2005}. When a non-zero neutrino magnetic moment (NMM) is present, the neutrino energy loss can increase due to the additional electromagnetic contribution to the neutrino emissivity \citep{Mori2020}. This additional energy loss, induced by physics transcending the Standard Model, can alter the rate of neutrino production, potentially impacting the star's structure \citep{Heger2008}. Based on GEMMA experiment \citep{Beda2013}, the current best experimental limit $\mu_{\nu}<2.9\times10^{-11}\mu_{B}$, where $\mu_{B} = e/2m_{\mathrm{e}}$ is the Bohr magneton, $\mathrm{e}=1.602\times10^{-19}\ \mathrm{C}$ is the electron charge and $m_{\mathrm{e}}=9.109\times10^{31}\ \mathrm{Kg}$ is the electron mass.

In this paper, we plan to investigate the effects of rotation and metallicity on the neutrino luminosity of massive stars, along with the influence of the NMM on stellar structure and evolution. We aim to explore the neutrino luminosity of stars with masses ranging from 20 to 90 $\rm M_{\odot}$, from the MS to Fe core collapse (FeCC) stages. Section 2 gives the details of software instruments, input physics, reaction networks and model selection. In section 3 we discuss the variation of our results in the luminosity of neutrinos with different solar masses, different metallicity, rotations, and the effect of NMM on the evolution of massive stars. The conclusions are given in section 4.

\section{MODEL}
We employ the open-source stellar evolution code (\texttt{MESA}, version 10398, \citealt{paxton2011, paxton2013, paxton2015, paxton2018, paxton2019}) to construct 20 to 90 $\rm M_{\odot}$ stellar models. Each stellar model is evolved from the Zero Age Main Sequence (ZAMS) until FeCC.
Following \cite{Farag2020}, we assume FeCC occurring when any location inside the stellar model reaches an infall velocity of 300 $\rm km \cdot s^{-1}$. The neutrino normalization $L_{\nu, \odot}=0.02398 \cdot L_{\gamma, \odot}=9.1795 \times 10^{31} \mathrm{erg\ s}^{-1}$ is adopted in \citet{Farag2020}.

\subsection{mass loss}
The mass-loss rates employed in the present paper are consistent with those used by \citet{Brott2011}. For stars hotter than approximately 25 $\rm kK$ with a surface hydrogen mass fraction $X_{\rm{S}}$ > 0.7, we utilize the wind recipe proposed by \citet{Vink2001}. For stars with low surface hydrogen mass fraction ($X_{\rm{S}}$ <0.4), we apply the mass-loss recipe for Wolf-Rayet (WR) stars proposed by \citet{Hamann1995}, scaled down by a factor of ten. The mass-loss rate is obtained through a linear interpolation between the mass-loss prescriptions provided by \citet{Vink2001} and \citet{Hamann1995} for hot stars with $X_{\mathrm{S}}$ between 0.4 and 0.7.

\subsection{convection}

We use MLT++ to treat convection, this prescription can calculate models of massive stars up to core collapse \citep{paxton2013}. Convection boundaries are calculated by the Ledoux criterion but not Schwarzschild criterion, mixing-length parameter of 1.5 is adopted \citep{Böhm-Vitense1958, Langer1991, Lü2017, Cui2018}. Semi-convection is a time-dependent diffusion process, the diffusion coefficient is given by \citet{Langer1983} as follows:

\begin{align}
    &&& D_{\mathrm{sc}}=\alpha_{\mathrm{sc}}\left(\frac{\mathrm{K}}{6 C_{\mathrm{P}} \rho}\right) \frac{\nabla_{\mathrm{T}}-\nabla_{\mathrm{ad}}}{\nabla_{\mathrm{L}}-\nabla_{\mathrm{T}}}
\end{align}

where $\nabla_{\mathrm{T}}$ is temperature gradient, $\nabla_{\mathrm{ad}}$ is adiabatic temperature gradient, K is radiative conductivity, $C_{\mathrm{P}}$ is specific heat at constant pressure, $\alpha_{\mathrm{sc}}$ is a dimensionless constant. In our work, we set $\rm \alpha_{sc}=0.01$ \citep{Wellstein2001}. The overshoot area is calculated by the step-function overshooting which depends on the overshooting parameter $\alpha_{\rm ov}$. Following \citet{Brott2011}, $\alpha_{\rm ov}=0.335$ in our simulations.

\subsection{neutrino emission}

The production of neutrinos inside stars can be roughly divided into two processes. One is weak nuclear processes, including $\beta^{\pm}$ decay and $e^{\pm}$ capture \citep{Kato2020}:

\begin{align}
	&&& \rm nuclei\ EC\ :\ (Z, A)+e^{-} \longrightarrow(Z-1, A)+\nu_{e} \\
	&&& \rm freep\ EC\ :\ p+e^{-} \longrightarrow n+\nu_{e} \\
	&&& \rm PC\ :\ (Z, A)+e^{+} \longrightarrow(Z+1, A)+\bar{\nu}_{e} \\
	&&& \rm \beta^+\ :\ (Z, A) \longrightarrow(Z-1, A)+e^{+}+\nu_{e} \\
	&&& \rm \beta^-\ :\ (Z, A) \longrightarrow(Z+1, A)+e^{-}+\bar{\nu}_{e}
\end{align}

such as $\rm ^{13}N(,e^+\nu_{e})^{13}C$, $\rm ^{15}O(,e^+\nu_{e})^{15}N$, $\rm ^{17}F(,e^+\nu_{e})^{17}O$ in CNO cycle, $\rm ^{18}F(,e^+\nu_{e})^{18}O$ during helium burning, $\rm ^{25}Al(,e^+\nu_{e})^{25}Mg$, $\rm ^{26}Al(,e^+\nu_{e})^{26}Mg$ during C burning, especially isotopes with $\rm A \approx 50 - 60$ are major contributor for neutrino luminosity during $\rm \beta$ processes, such as $\rm ^{51,52,53}V$, $\rm^{53,55,63}Cr$, $\rm^{54,55,56}Mn$, $\rm^{53,54,55}Fe$, $\rm^{55,56,57}Co$ \citep{Patton2017}. For the massive star model, the large nuclear reaction network can better track the neutrinos produced by each nuclear reaction process, so we select mesa\_204.net as the nuclear reaction network.

The other is thermal processes, thermal neutrino energy losses are from \citep{Itoh1996}, including:
\vspace{0.2cm} 

\noindent pair annihilation$\rm \left(T>10^9\ K\right)$ 
\begin{align}
	&&& e^{+}+e^{-} \longrightarrow \nu+\bar{\nu}
\end{align}
photon-neutrino$\rm \left(T<4\times10^8\ K, \rho<10^5\ g\ cm^{-3}\right)$ 
\begin{align}
	&&& e^{-}+\gamma \longrightarrow e^{-}+\nu+\bar{\nu}
\end{align}
plasma decay$\rm \left(10^7<T<10^8\ K, 10^4<\rho<10^7\ g\ cm^{-3}\right)$ 
\begin{align}
	&&& \rm \gamma^{*} \longrightarrow \nu+\bar{\nu}
\end{align}
bremsstrahlung$\rm \left(10^8<\rho<10^{10}\ g\ cm^{-3}\right)$
\begin{align}
	&&& e^{-}+(Z, A) \longrightarrow e^{-}+(Z, A)+\nu+\bar{\nu}
\end{align} 
The MESA nuclear reaction rates are a combination of rates from NACRE \citep{Angulo1999} and JINA REACLIB \citep{Cyburt2010}. Treatment of screening corrections are from \citep{Chugunov2007}, which enhance nuclear reactions rates in dense plasmas. All the weak reaction rates are based on \citep{Langanke2000, Oda1994, Fuller1985}.  

\subsection{Rotation and metallicity}
Rotation and metallicity have great effects on the massive stars. Rapid rotation may trigger some instabilities in the stellar internal, which can result in chemical homogenous evolution \citep{Maeder1980, Martins2009, Martins2013, Hainich2015, Schootemeijer2018}.
Following \citet{Heger2000}, we take into account several instabilities induced by rotation that result in mixing: Eddington-Sweet circulation instability, dynamical and secular shear instability, and the Goldreich-Schubert-Fricke instability. Additionally, the transport of angular momentum facilitated by magnetic fields via the Spruit-Tayler dynamo mechanism is also taken into consideration \citep{Brott2011}.
According to \citet{Heger2000}, the efficiency of rotational mixing, $f_{\mathrm{c}}$, is considered as 1/30. The inhibition parameter $f_{\mathrm{\mu}}$ of chemical gradient on the efficiency of rotational mixing process is set to 0.1 \citep{Yoon2006}.
In order to discuss the effects of rotation, we take $\omega/\omega_{\rm crit}$=0 and 0.7 for different simulations,
where $\omega/\omega_{\rm crit}$ is the ratio of rotational angular velocity to critical rotational angular velocity.
Simultaneously, metallicity can affect the efficiency of chemical homogenous evolution. Therefore, we select $Z=0.014$ for the high metallicity,
0.001 and 0.0001 for metal-poor stars.

In addition, the mass-loss rate of rotating massive stars would be enhanced by
\begin{align}
    &&&	\dot{M}=\left(\frac{1}{1-\omega / \omega_{\text {crit }}}\right)^{\beta} \dot{M}_{v_{\mathrm{rot}}=0}
\end{align}
where $\rm \beta=0.43$ \citep{Langer1998}.

\subsection{neutrino magnetic moment}

Following \cite{Mori2020}, the additional energy loss induced by NMM has two ways: plasmon decay and neutrino pair production. The energy loss rate due to the plasmon decay is given by
\begin{align}
	&&& \epsilon_{\text {plas }}^\mu=0.318\left(\frac{\omega_{\mathrm{pl}}}{10 \mathrm{keV}}\right)^{-2}\left(\frac{\mu_v}{10^{-12} \mu_{\mathrm{B}}}\right)^2 \epsilon_{\mathrm{plas}}
\end{align}
where $\rm \epsilon_{\text {plas }}$ is the standard plasmon decay rate \citep{Itoh1996} and  $\rm \omega_{\text {pl }}$ is the plasma frequency \citep{Raffelt1996}
\begin{align}
	&&& \omega_{\mathrm{pl}}=28.7 \mathrm{eV} \frac{\left(Y_{\mathrm{e}} \rho\right)^{\frac{1}{2}}}{\left(1+\left(1.019 \times 10^{-6} Y_{\mathrm{e}} \rho\right)^{\frac{2}{3}}\right)^{\frac{1}{4}}}
\end{align}
where $Y_{\mathrm{e}}$ represents the electron mole fraction, $\rm \rho$ denotes the density in units of $\rm g\ cm^{-3}$. The energy loss rate due to the pair production is given by \citep{Heger2008}
\begin{align}
	&&& \epsilon_{\text {pair }}^\mu=1.6 \times 10^{11} \mathrm{erg}^{-1} \mathrm{~s}^{-1}\left(\frac{\mu_v}{10^{-10} \mu_{\mathrm{B}}}\right)^2 \frac{e^{-\frac{118.5}{T_8}}}{\rho_4}
\end{align}
where $\rm T_{8}$ is the temperature in units of $\rm 10^{8}\ K$ and $\rm \rho_{4} =\rho/(10^{4} g\ cm^{-3})$. We added these two additional energy loss calculation formulas to $\rm run\_star\_extras$.

\section{Result }
Using MESA, we simulate the structure and evolution of
8 massive star models including different initial masses from 20 to 90 M$_\odot$ with a mass interval of 10 M$_\odot$, different initial rotational
velocities ($\omega/\omega_{\rm crit}=0$ and 0.7), and different metallicities (Z=0.014, 0.001 and 0.0001).
Following \cite{Farag2020}, we employ the neutrino H-R diagram to present the neutrino luminosities throughout the entire life cycles of these massive stars,
and discuss how stellar rotational velocity and metallicity affect stellar neutrino luminosity.

\subsection{H-R diagram and neutrino H-R diagram of massive star}
\label{subsection:3.1}
\begin{figure*}
    \centering
        \includegraphics[width=0.8\textwidth,height=0.8\textwidth]{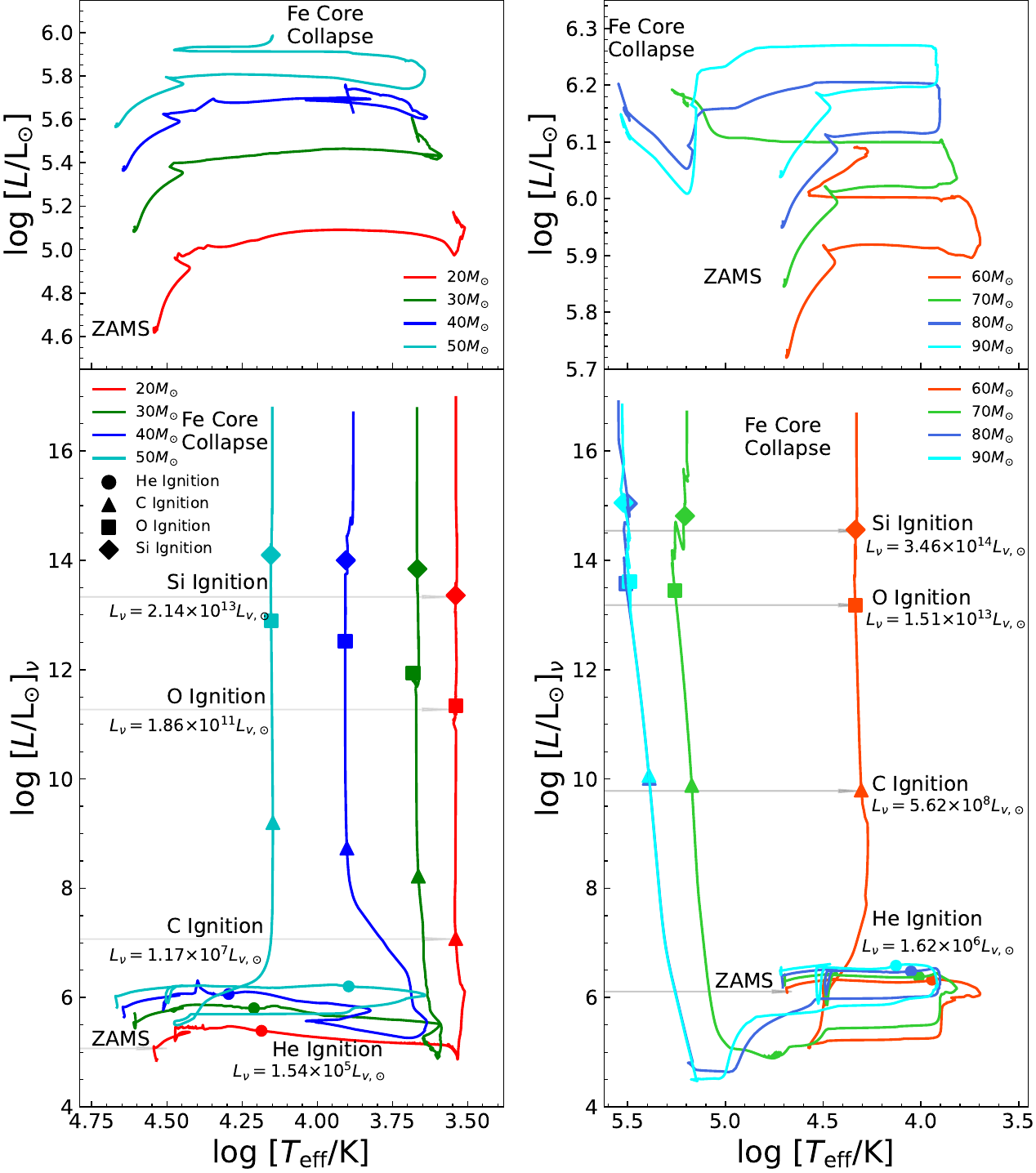}  
  
        \caption{The evolution of 20 $\rm M_{\odot}$ to 90 $\rm M_{\odot}$ star with non-rotation and high metallicity ($Z=0.014$) in the H-R diagram (top panels) and neutrino H-R diagram (bottom panels), the left is 20 $\rm M_{\odot}$ to 50 $\rm M_{\odot}$ star, right is 60 $\rm M_{\odot}$ to 90 $\rm M_{\odot}$ star. The positions of the neutrino luminosity at the ignition of He, C, O, and Si are given.}
        \label{fig:1}
    \end{figure*}

Figure \ref{fig:1} shows the 20 to 90 M$_{\odot}$ stellar evolutionary tracks in both the H-R diagram and the neutrino H-R diagram.
  It clearly illustrates that the neutrino luminosity during the MS phase increases correspondingly as the stellar mass increases.
  In massive stars, the primary source of neutrinos is the Carbon-Nitrogen-Oxygen (CNO) cycle,
  which is highly temperature-sensitive ($\varepsilon_{\rm CNO} \propto T^{17}$) \citep{Bahcall1989}.
  As a result, nuclear reactions are more intense in more massive stars, which leads to higher neutrino luminosity.

  As shown in Figure \ref{fig:1}, the neutrino luminosity of the star decreases steadily during the helium burning phase. During this phase, only one weak process, namely $\rm ^{18}F(,e^+\nu_{e})^{18}O$, generates neutrinos in the 3$\alpha$ reaction \citep{Odrzywolek2010}. Consequently, the neutrino luminosity during the helium burning phase arises primarily from the burning of the hydrogen shell. At this stage, the high mass loss rate leads to a reduction in the hydrogen envelope, a decrease in the hydrogen shell luminosity, and a corresponding decrease in the neutrino luminosity. For more massive stars, this phenomenon is even more pronounced due to greater mass loss rate.

Then, the neutrino luminosity experiences a rapid increase following carbon burning. Figure \ref{fig:1} also marks the neutrino luminosity of 20 $\rm M_{\odot}$ and 60 $\rm M_{\odot}$ stars when helium depletion and durning carbon, oxygen, and silicon burning. When the Carbon is ignited, the internal temperature of the massive star exceeds $8\times10^8\rm\ K$, leading to the domination of thermal process, particularly electron-positron pair annihilation, in neutrino production. At this stage, massive stars are often called neutrino-cooled stars \citep{Odrzywolek2004}. This is due to the trapping of photons as the density inside the star increases, while neutrinos can freely escape with their extremely small cross sections. In addition, neutrinos carry away a significant amount of energy during C, Ne, O, and Si burning, which drastically reduces the lifetime of nuclear burning. As the temperature inside the star increases rapidly, the thermal neutrino energy loss becomes more significant. Therefore, the sensitivity of neutrino energy loss to temperature allows massive stars to evolve more rapidly towards core collapse.

\begin{figure*}
    \centering
        \includegraphics[width=0.8\textwidth,height=\textwidth]{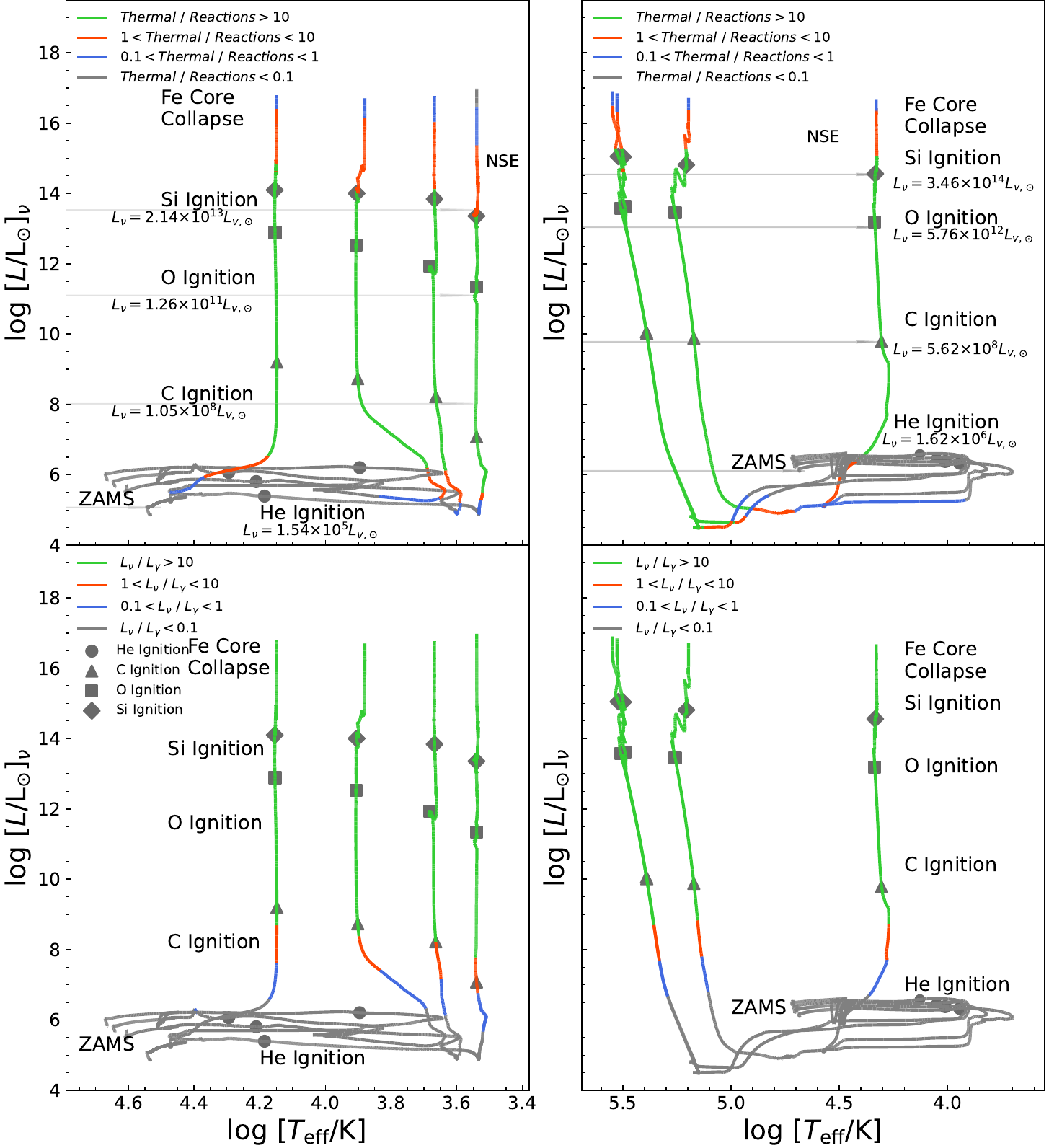}

        \caption{The neutrino luminosity produced by the nuclear reaction and thermal processes (the top two panels).
                The colorful solid lines are the ratio of the nuclear reaction neutrino luminosity to the thermal
                neutrino luminosity plotted along stellar evolution tracks in a neutrino H-R
                diagram. Gray curves indicate where nuclear reaction neutrinos dominate,
                green curves where thermal neutrinos dominate, and blue and red curves where the
                reaction and thermal neutrino luminosities are within a factor of 10.
                The bottom two panels show the ratio of the neutrino luminosity to the photon luminosity during different evolutionary stages.
                The colorful solid lines give the ratio
                of the neutrino luminosity to the photon luminosity in a neutrino H-R diagram.
                The left panels are for 20 to 50 $\rm M_{\odot}$ models with non-rotation and high metallicity (($Z=0.014$)), the right panels are for 60 to 90 $\rm M_{\odot}$ models with non-rotation and solar metallicity. The positions of the neutrino luminosity at the ignition of He, C, O, and Si are given.
        }
        \label{fig:2}
    \end{figure*}

As mentioned in the above section, both nuclear reaction and thermal process can produce neutrinos.
The upper two panels of Figure \ref{fig:2} depict the ratio of nuclear reaction neutrino luminosity to thermal process neutrino luminosity, traced along the stellar evolution trajectories in a neutrino H-R diagram. The lower panels of Figure \ref{fig:2} illustrate the ratio of neutrino luminosity to photon luminosity, also traced along the stellar evolution trajectories in a neutrino H-R diagram. During hydrogen and helium burning, the majority of neutrinos are generated via nuclear reactions, and massive stars primarily radiate their energy through photons. Upon the commencement of carbon burning, the dominant source of neutrinos transitions from nuclear reactions to thermal processes. Consequently, neutrinos become the principal conduit for energy loss, and the neutrino luminosity significantly surpasses the photon luminosity.

\begin{figure}
	\includegraphics[width=\columnwidth]{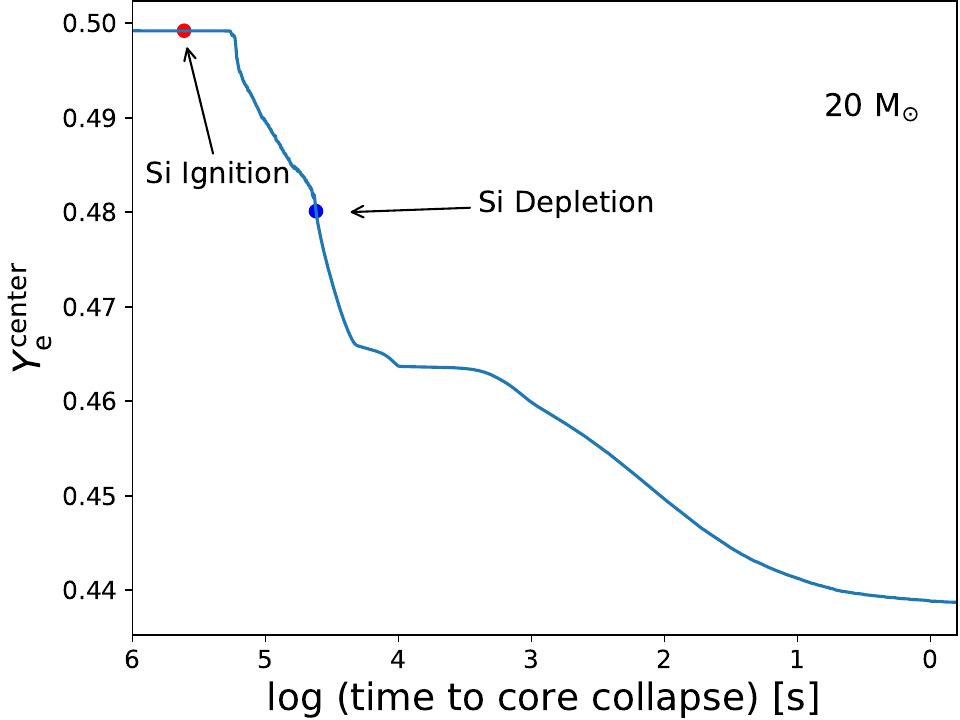}
    \vspace{-0.6cm}
    \caption{Evolution of $Y_{e}$ in the center of a 20 $\rm M_{\odot}$ star with non-rotation and high metallicity ($Z=0.014$) after Oxygen depletion. The red dot represents Si ignition, the blue dot represents Si depletion.}
    \label{fig:3}
\end{figure}

In particular, the neutrino luminosity produced by the nuclear reactions becomes gradually important
during Si burning. The main reason is given in Figure \ref{fig:3}, which illustrates the change of central electron
mole number $Y_{\mathrm{e}}$ in a 20 $\rm M_{\odot}$ star with non-rotation and high metallicity ($Z=0.014$) after oxygen depletion. It shows that the number of electrons in the star decreases during silicon burning. This reduction can be attributed to an increase in the internal density and temperature,
which leads to an increase in the electron degeneracy pressure. As a result, the mean energy of electrons becomes
larger than that of many heavy nuclei, increasing the rate of $\rm e^{-}$ capture processes in the stellar interior.
Once the core temperature of the massive star reaches $\rm 5\times10^9\ K$, nuclear statistical equilibrium (NSE) is reached,
and the $\rm e^{-}$ capture process is significantly enhanced \citep{Kato2017}. During the NSE phase, $\beta$ decay
and electron capture (EC) on heavy nuclei dominate the production of neutrinos. As demonstrated in the top rows of Figure \ref{fig:2},
once the core begins to collapse, a large number of $\rm e^{-}$ captures enable nuclear reaction processes to produce far more neutrinos than thermal processes.

\subsection{Effect of metallicity on neutrino luminosity}
\begin{figure}
	\includegraphics[width=\columnwidth,height=0.3\textheight]{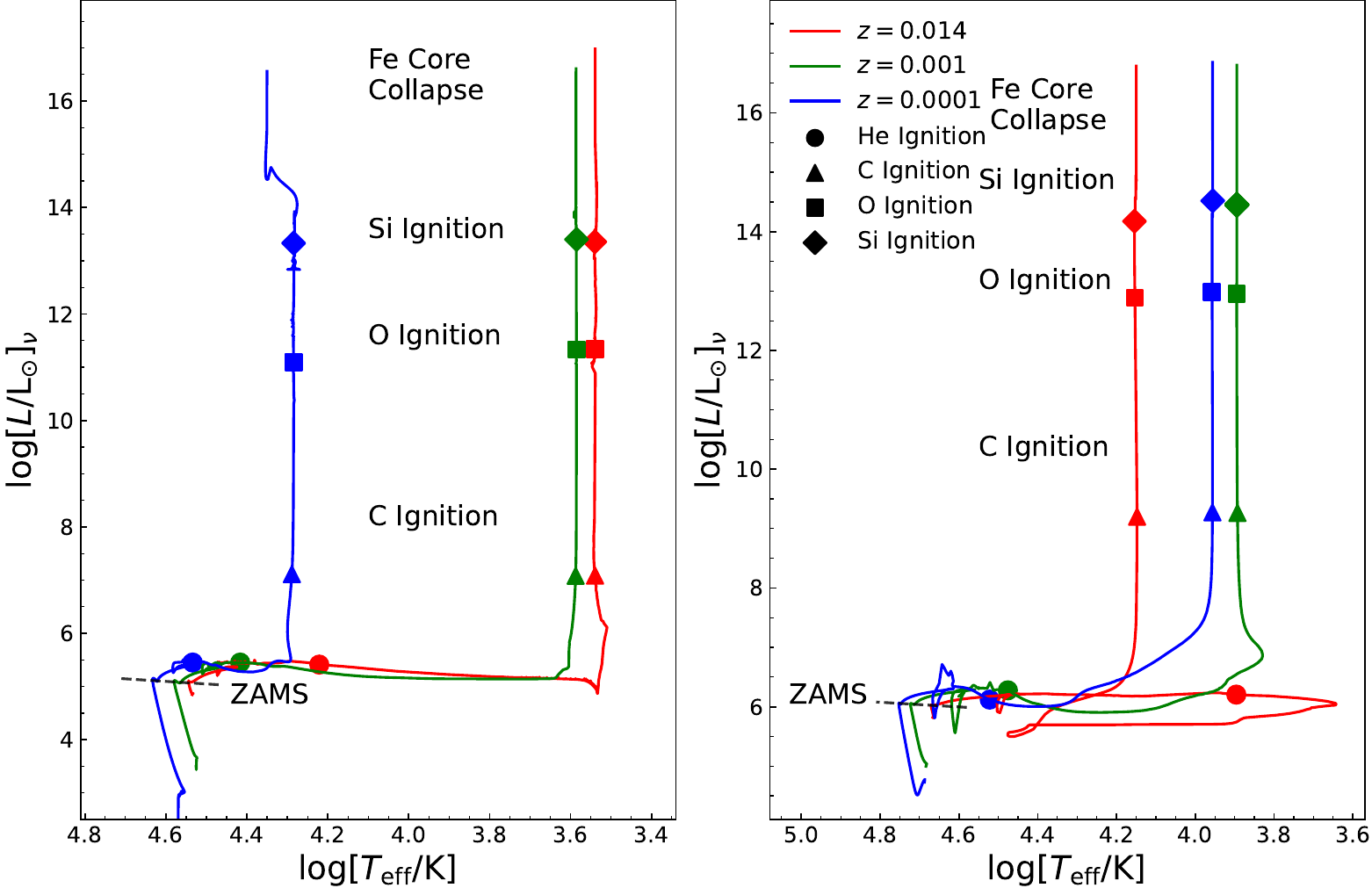}
    \vspace{-0.6cm}
    \caption{The neutrino H-R diagram. The left panel is 20 $\rm M_{\odot}$ star with non-rotation, right panel is 50 $\rm M_{\odot}$ star with non-rotation. The red line is $\rm Z=0.014$ model, green line is $\rm Z=0.001$ model, blue line is $\rm Z=0.0001$ model. 
    The ZAMS is located on the dashed line. The positions of the neutrino luminosity at the ignition of He, C, O, and Si are given.}
    \label{fig:4}
\end{figure}

In order to discuss its effects on neutrino luminosity, we present Figure \ref{fig:4}, which shows the neutrino H-R diagram of massive stars with different metallicities. In this section, we do not take into account the effect of rotation on massive stars.
For the low-metallicity models, the neutrino luminosity is higher than that of models with high metallicity ($Z=0.014$)
during MS because they have a higher central temperature.
Simultaneously, the low-metallicity model has a weaker mass-loss rate, which results in much hydrogen shell, and heavier helium core.
Therefore, the massive stars with a low metallicity have higher neutrino luminosity during helium burning.

\subsection{Effect of rotation on neutrino luminosity}
Rotation is widely acknowledged to profoundly influence stellar structure and evolution. 
Here, taking the models with 20, 50 and 90 M$_\odot$ as examples, we discuss its effects on the neutrino luminosity. 

\begin{figure*}
	\includegraphics[width=0.9\textwidth]{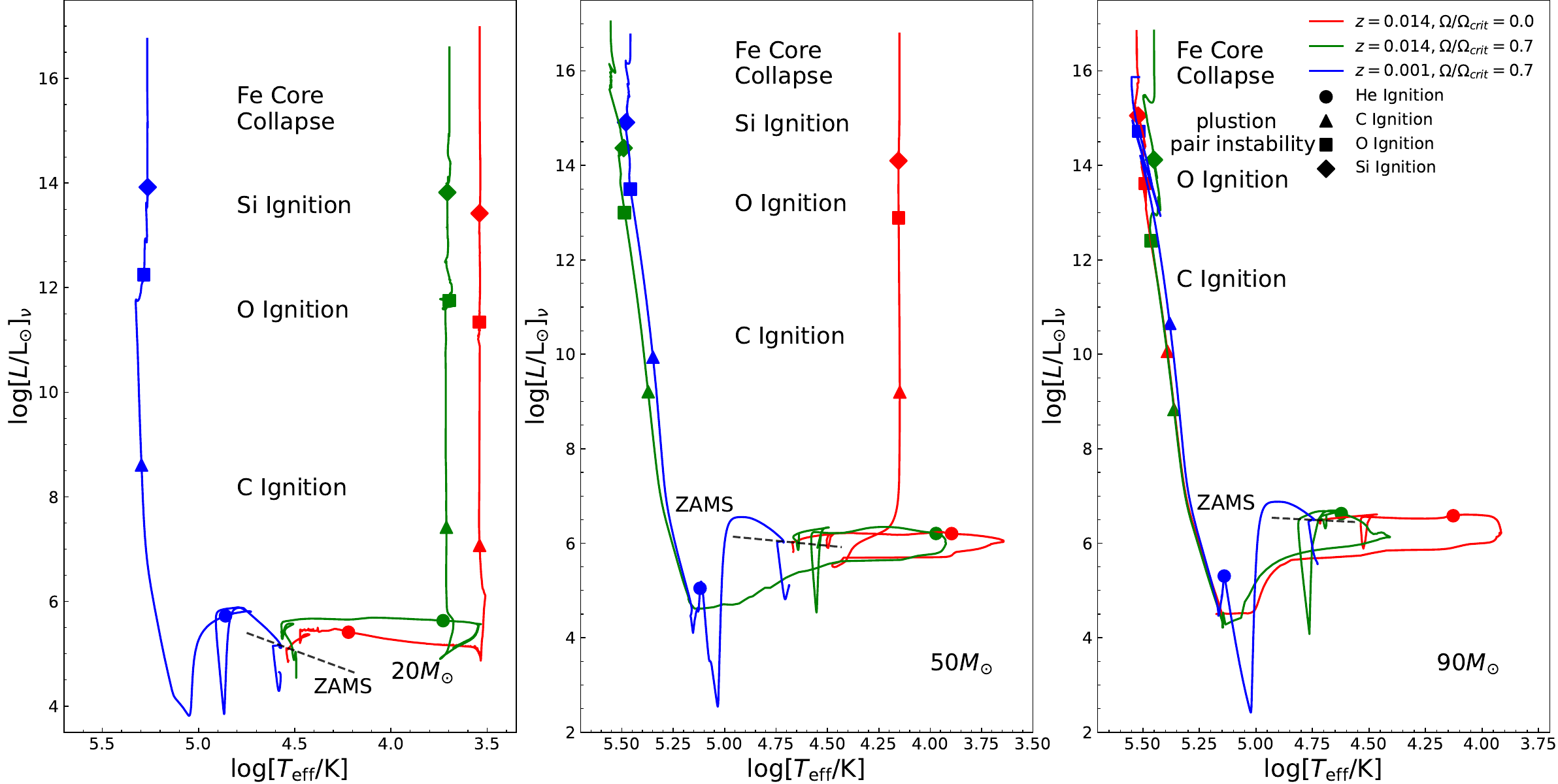}

    \caption{The neutrino H-R diagram. From left to right is 20, 50 and 90 $\rm M_{\odot}$ star, respectively. 
    The red line is high metallicity ($Z=0.014$) without rotation model,
    green line is high metallicity with high rotation ($\rm \omega/\omega_{crit}=0.7$) model, blue line is $\rm Z=0.001$ with high rotation model.
    The ZAMS is located on the dashed line. The positions of the neutrino luminosity at the ignition of He, C, O, and Si are given.}
    \label{fig:5}
\end{figure*}

\begin{figure*}
	\includegraphics[width=\textwidth]{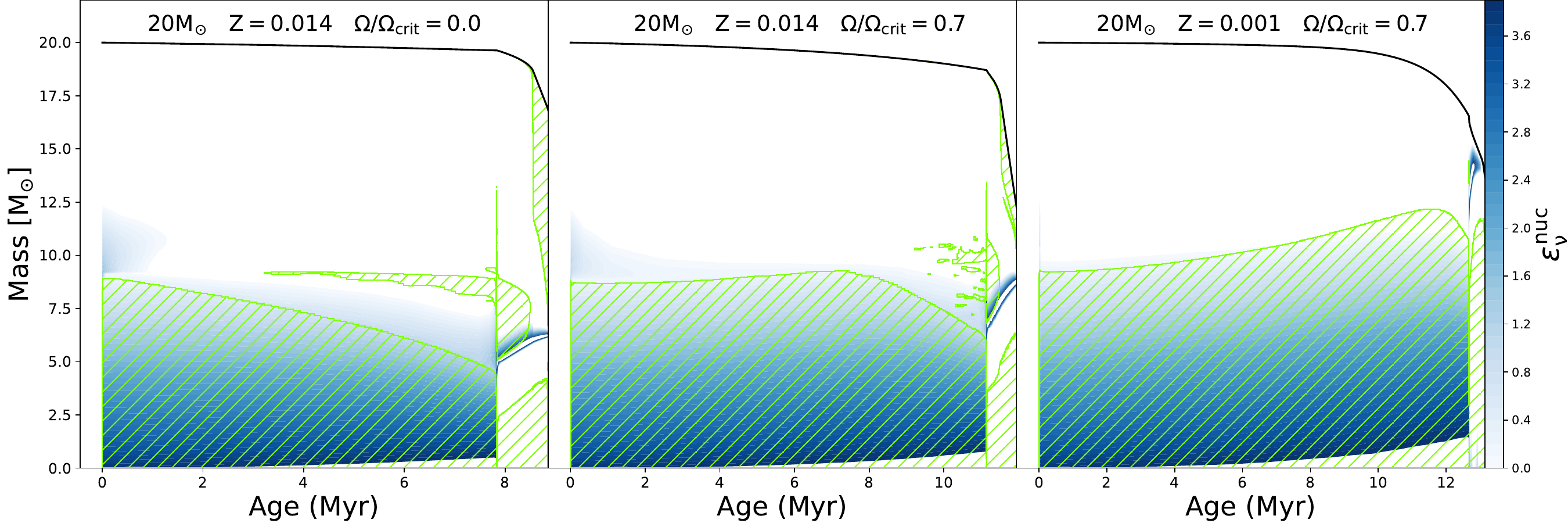}
    \caption{The Kippenhahn diagrams of 20 $\rm M_{\odot}$ star. The green shaded areas indicate convection zones, colors indicate the specific energy rate due to neutrino emission $\rm \epsilon_{\nu}$. From left to right are $\rm Z=0.014$ and $\omega/\omega_{\mathrm{crit}}=0$, $\rm Z=0.014$ and $\omega/\omega_{\mathrm{crit}}=0.7$, $\rm Z=0.001$ and $\omega/\omega_{\mathrm{crit}}=0.7$ models.}
    \label{fig:6}
\end{figure*}

Figure \ref{fig:5} displays 20, 50 and 90 $\rm M_{\odot}$ stellar evolutionary tracks with different rotation and metallicity in the neutrino H-R diagram. At high metallicity ($Z=0.014$), it is obvious that the neutrino luminosity increases during MS for 20 $\rm M_{\odot}$ star with high rotation ($\omega/\omega_{\mathrm{crit}}=0.7$). As described by \citet{Georgy2013}, rotation induces mixing and brings light elements, like hydrogen, into the core. This leads to larger convective and burning regions during MS, as shown in middle panel of Figure \ref{fig:6}, which increases the MS lifetime and enhances the energy yield of the CNO cycle. For massive stars, neutrinos are mainly produced by the CNO cycle during hydrogen burning. Therefore, the more efficient the CNO cycle inside the star, the higher the neutrino luminosity during hydrogen burning. The high rotation model ($\omega/\omega_{\mathrm{crit}}=0.7$) with high metallicity ($Z=0.014$) has a 0.4 dex larger neutrino luminosity than the non-rotating model during hydrogen burning.
At low metallicity, the surface angular momentum loss is reduced due to diminished stellar wind losses. In our simulations, the 20 M$_\odot$ model with rapid rotation ($\omega/\omega_{\mathrm{crit}}=0.7$) can undergo a strong mixing process and result in efficient chemically homogeneous evolution during the MS phase. The right panel of Figure \ref{fig:6} illustrates that the convection and burning regions of the models with low metallicity and rapid rotation are
larger than models with high metallicity and rapid rotation. This results in a more efficient CNO cycle and, therefore, a higher neutrino luminosity.

Regarding the neutrino HR diagram of 50 and 90 $\rm M_{\odot}$ stars depicted in Figure \ref{fig:5}, at high metallicity ($Z=0.014$) and high rotation ($\omega/\omega_{\mathrm{crit}}=0.7$), these stars can undergo intense star wind losses and angular momentum loss during its MS lifetime. Therefore, despite the high rotation rate, mixing is not efficient, and there is no significant increase in neutrino luminosity compared to a 20 $\rm M_{\odot}$ star. 
At low metallicity and high rotation, the evolution of 50 and 90 $\rm M_{\odot}$ stars follow a trajectory similar to that of a 20 $\rm M_{\odot}$ star. The smaller mass loss and more massive core allow these stars also undergo quasi-chemical homogeneous evolution. Therefore, the neutrino luminosity of the 50 and 90 $\rm M_{\odot}$ stars increases significantly during the MS. In fact, the rotation rate required for 50 and 90 $\rm M_{\odot}$ star to achieve chemically homogeneous evolution is smaller than that for a 20 $\rm M_{\odot}$ star \citep{Yoon2006}. 

During the helium burning, neutrino H-R diagram of Figure \ref{fig:5} displays the high rotation ($\omega/\omega_{\mathrm{crit}}=0.7$) models  show more significant drops in neutrino luminosity than non-rotation. The drop occurs because mixing processes take away significant amounts of hydrogen from the stellar surface, leaving a hydrogen shell that is too thin to produce more neutrinos. This phenomenon is particularly pronounced in low-metallicity models, as the star undergoes quasi-chemical homogeneous evolution and completely loses its hydrogen shell, causing the neutrino luminosity to drop almost linearly during the helium burning. Simultaneously, during the helium burning, masstive stars with high rotation lose more hydrogen shell. For more massive stars, such as our 50 and 90 $\rm M_{\odot}$ star models, their helium cores can even be eroded by intense star wind losses.

\begin{figure}
	\includegraphics[width=\columnwidth]{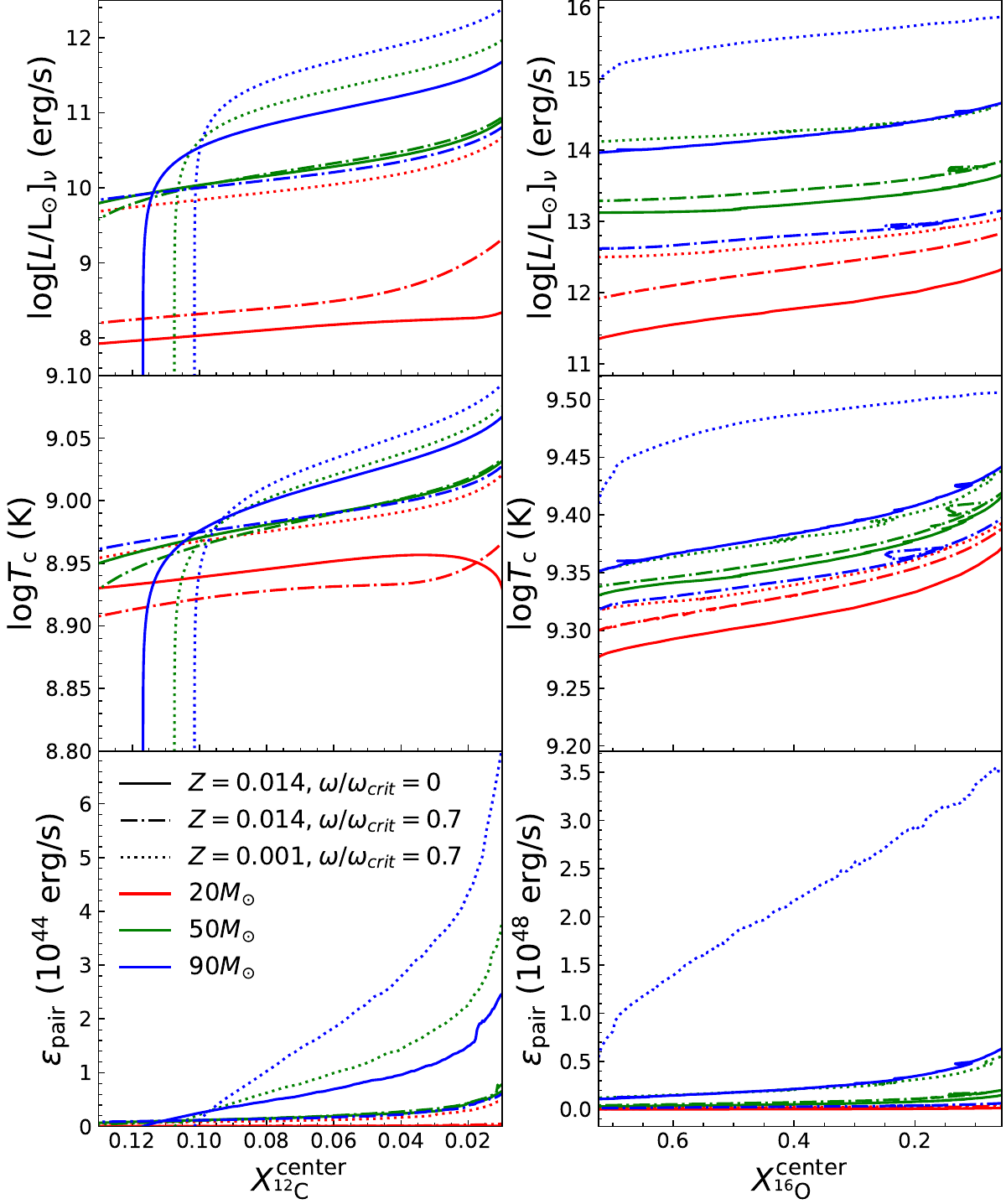}
    \vspace{-0.6cm}
    \caption{The neutrino luminosity, center temperature($T_{c}$) and energy yield of electron pair annihilation (from top to bottom) during C (left) and O (right) burning. 
     The solid line is $\rm Z=0.014$ and $\omega/\omega_{\mathrm{crit}}=0$ star model, the dashdot line is $\rm Z=0.014$ and $\omega/\omega_{\mathrm{crit}}=0.7$ star model, the dotted line is $\rm Z=0.001$ and $\omega/\omega_{\mathrm{crit}}=0.7$ star model. The red line is 20 $\rm M_{\odot}$, the green line is 50 $\rm M_{\odot}$, the blue line is 90 $\rm M_{\odot}$.}    
    \label{fig:7}
\end{figure}
Rotation and metallicity can also affect the neutrino luminosity at the advanced burning stage of massive stars. The top panel of Figure \ref{fig:7} shows the neutrino luminosity of 20 $\rm M_{\odot}$, 50 $\rm M_{\odot}$, and 90 $\rm M_{\odot}$ stars with different rotation and metallicity during carbon and oxygen burning. For the 20 $\rm M_{\odot}$ star, high rotation can increase the neutrino luminosity, with this increase being more pronounced at low metallicity. This is because the 20 $\rm M_{\odot}$ star with high rotation ($\omega/\omega_{\mathrm{crit}}=0.7$) can have a larger $\mathrm{He}$ and $\mathrm{CO}$ core mass than those without rotation, and the lower the metallicity, the larger the increase in mass. Additionally, massive stars with low metallicity show a higher temperature after helium burning, as shown in the middle panel of Figure \ref{fig:7}. In section \ref{subsection:3.1}, it was mentioned that the neutrino luminosity after helium burning to oxygen depletion is mainly determined by the thermal process, particularly pair annihilation. It is independent of density and sensitive to temperature, as indicated by \citet{Itoh1996}, so an increase in neutrino luminosity can occur significantly, as shown in the bottom panel of Figure \ref{fig:7}. These factors contribute to the increase in neutrino luminosity at later stages of stellar evolution. Although the $T_{\mathrm{c}}$ of a 20 $\rm M_{\odot}$ star with high rotation ($\omega/\omega_{\mathrm{crit}}=0.7$) is slightly lower than that of a non-rotating star during carbon burning, the larger CO core mass leads to a higher neutrino luminosity.

\begin{figure}
	\includegraphics[width=\columnwidth]{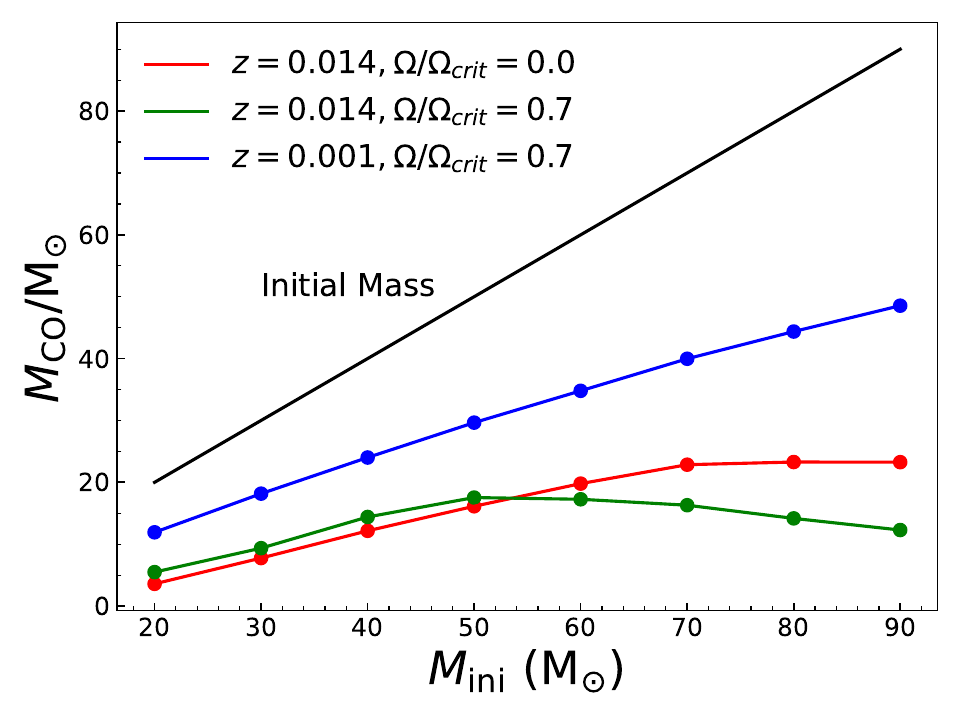}
    \vspace{-0.6cm}
    \caption{$\rm M_{CO}-M_{INI}$ relation. The dark line is initial mass.}    
    \label{fig:8}
\end{figure}

For the 50 $\rm M_{\odot}$ star with high rotation ($\omega/\omega_{\mathrm{crit}}=0.7$) and high metallicity ($Z=0.014$) in Figure \ref{fig:7}, the neutrino luminosity do not increase significantly during carbon and oxygen burning, similar to the MS. This is because as the initial mass of the massive star increases at high metallicity, the effect of rotation on $M_{\mathrm{CO}}$ gradually decreases, as shown in Figure \ref{fig:8}. For massive stars with $\rm M>50\ M_{\odot}$, the global effect of mass loss surpasses the influence of rotation, which even decreases the CO core mass \citep{limongi2018}. For example, as our 90 $\rm M_{\odot}$ star (which is shown by blue lines in Figure \ref{fig:7}) with high metallicity model, high rotation ($\omega/\omega_{crit}=0.7$) reduce the CO core mass by $32\%$ compared to non-rotation. As a result, the average neutrino luminosity is always reduced by about $60\%$ and $95\%$ during carbon and oxygen burning, respectively, compared to the 90 $\rm M_{\odot}$ non-rotation and high metallicity model. At low metallicity, the effect of mass loss can be neglected. Both 50 and 90 $\rm M_{\odot}$ star models with high rotation ($\omega/\omega_{\mathrm{crit}}=0.7$) have a larger CO core mass than non-rotating models after helium depletion, resulting in a significant increase in neutrino luminosity during carbon and oxygen burning.

It is worth noting that our 90 $\rm M_{\odot}$ star model with high rotation and low metallicity can enter the pulsation pair instability regime during O burning. As described by \citet{Woosley2002}, massive stars with a CO core mass exceeding 35 $\rm M_{\odot}$ have a structural adiabatic index $\Gamma_{1}$ below $4/3$, which can lead to instability. To avoid this instability, we stop the calculation as soon as the instability sets in. 

\begin{figure*}
	\includegraphics[width=0.75\textwidth,height=0.4\textheight]{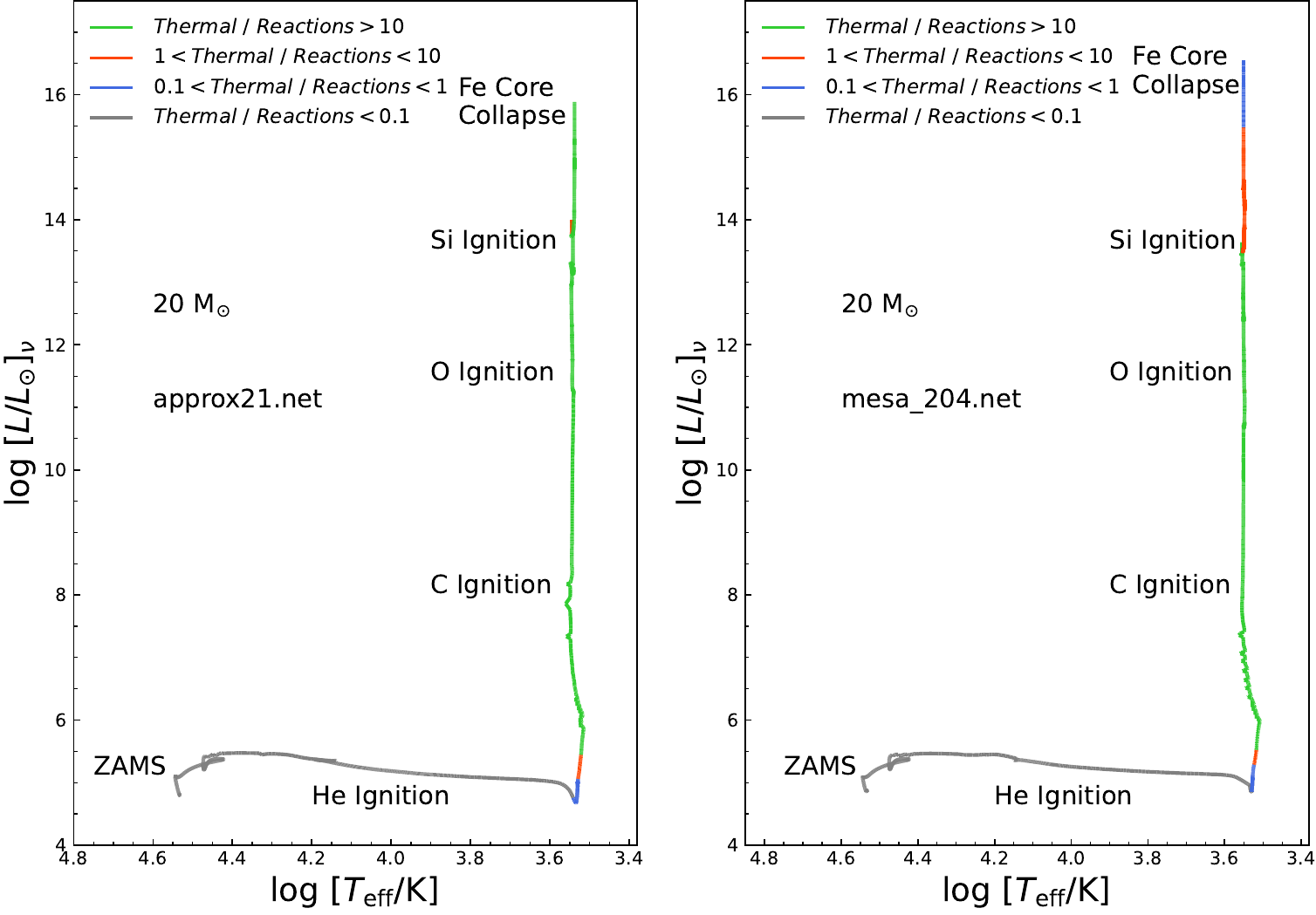}

    \caption{Similarly to Figure \ref{fig:2}, but for models with different network. The left panel is for  
    approx21.net, while the right panel is for $mesa\_204.net$, respectively.}    
    \label{fig:9}
\end{figure*}

\subsection{Effect of network on neutrino luminosity}
The selection of nuclear network plays an important role in the nuclear neutrino luminosity of massive stars. Figure \ref{fig:9} shows the ratio of thermal process to nuclear process neutrino luminosity along the stellar evolution track of 20 $\rm M_{\odot}$ star with non-rotation and high metallicity ($Z=0.014$) in a neutrino H-R diagram when adopted the basic network approx21.net and large network $mesa\_204.net$, respectively. Obviously, in a small network, electron capture and $\beta$ decay contributes little to the neutrino production due to the absence of a large number of isotopes, so that nuclear neutrinos are still much smaller than thermal neutrinos after silicon burning. In a larger network, however, it not only tracks a large number of nuclear neutrinos after Si burning, but also shows a higher neutrino luminosity.

\subsection{The effects of neutrino magnetic moments on the evolution of massive stars}
\begin{figure}
	\includegraphics[width=\columnwidth]{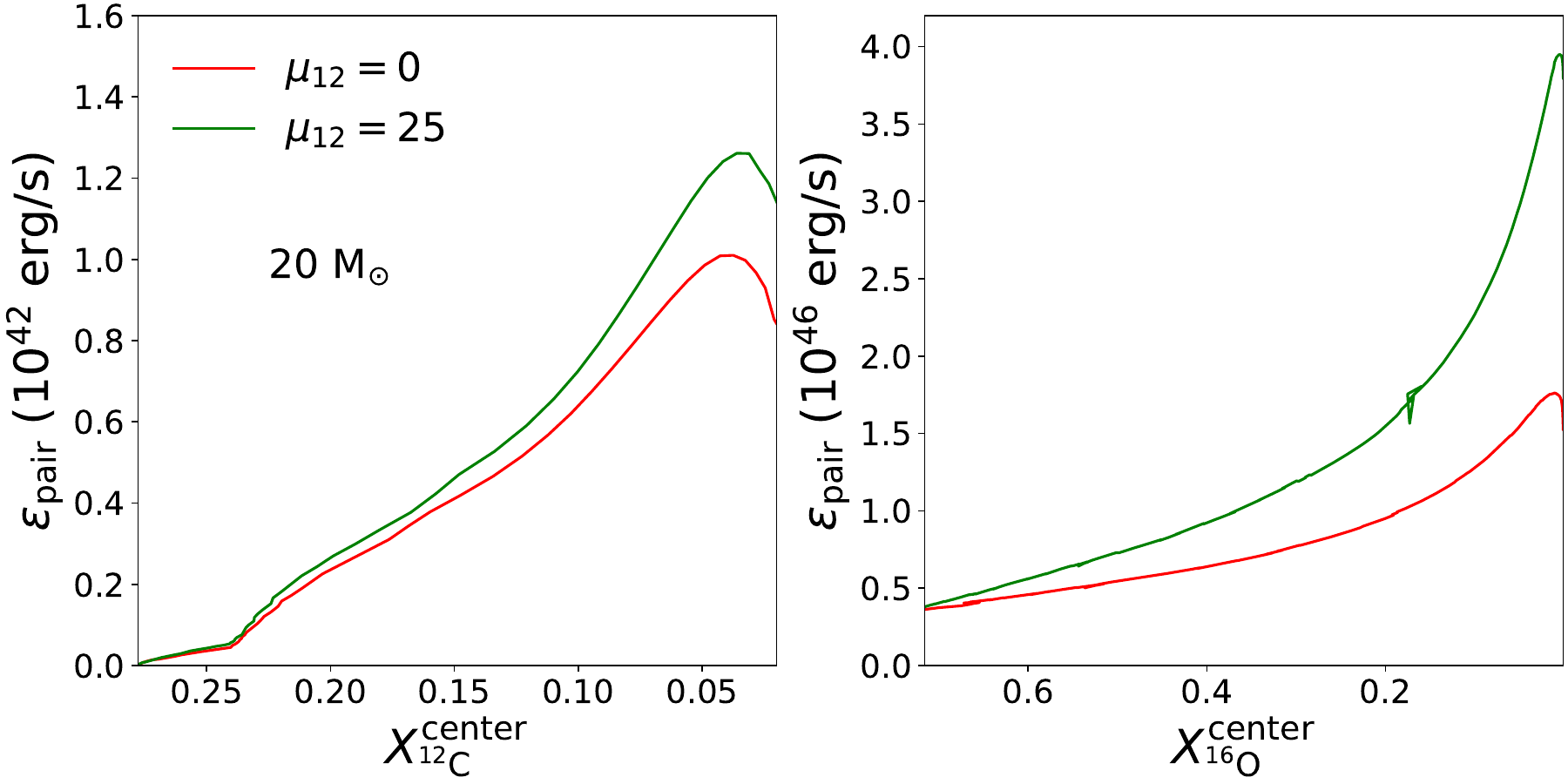}
    \vspace{-0.6cm}
    \caption{The energy yield of electron pair annihilation during C (left) and O (right) burning when the $\rm \mu_{-12}=0$ and $\rm \mu_{-12}=25$, respectively. 
     $\rm \mu_{-12}$ is the NMM in units of $\rm 10^{-12} \mu_{B}$, the star mass is 20 M$_{\odot}$.}    
    \label{fig:10}
\end{figure}

Based on the above simulations, massive stars produce a huge number of neutrinos during their lifetime. If neutrinos have a magnetic moment, their effects on the evolution of massive stars may be significant. It is because the NMM can cause additional energy loss when pair annihilation occurs inside the star, as shown in Figure \ref{fig:10}. The neutrino losses from pair annihilation dominate the evolution of massive stars from He burning to oxygen depletion. Therefore, we have evolved all the models in this paper with $\rm \mu_{-12}=25$, where $\rm \mu_{-12}$ is the NMM in units of $\rm 10^{-12} \mu_{B}$.

When $\rm \mu_{-12}=25$, the effects of NMM on the evolution of massive stars can be negligible. The energy loss caused by such a small NMM cannot change stellar destiny of massive stars from ZAMS to FeCC. This result is consistent with that of \citet{Heger2008}, who examined the effect of NMM on stellar evolution in the mass range of $7\ \mathrm{M_{\odot}} \lesssim M \lesssim 18\ \mathrm{M_{\odot}}$.

\begin{figure}
	\includegraphics[width=\columnwidth]{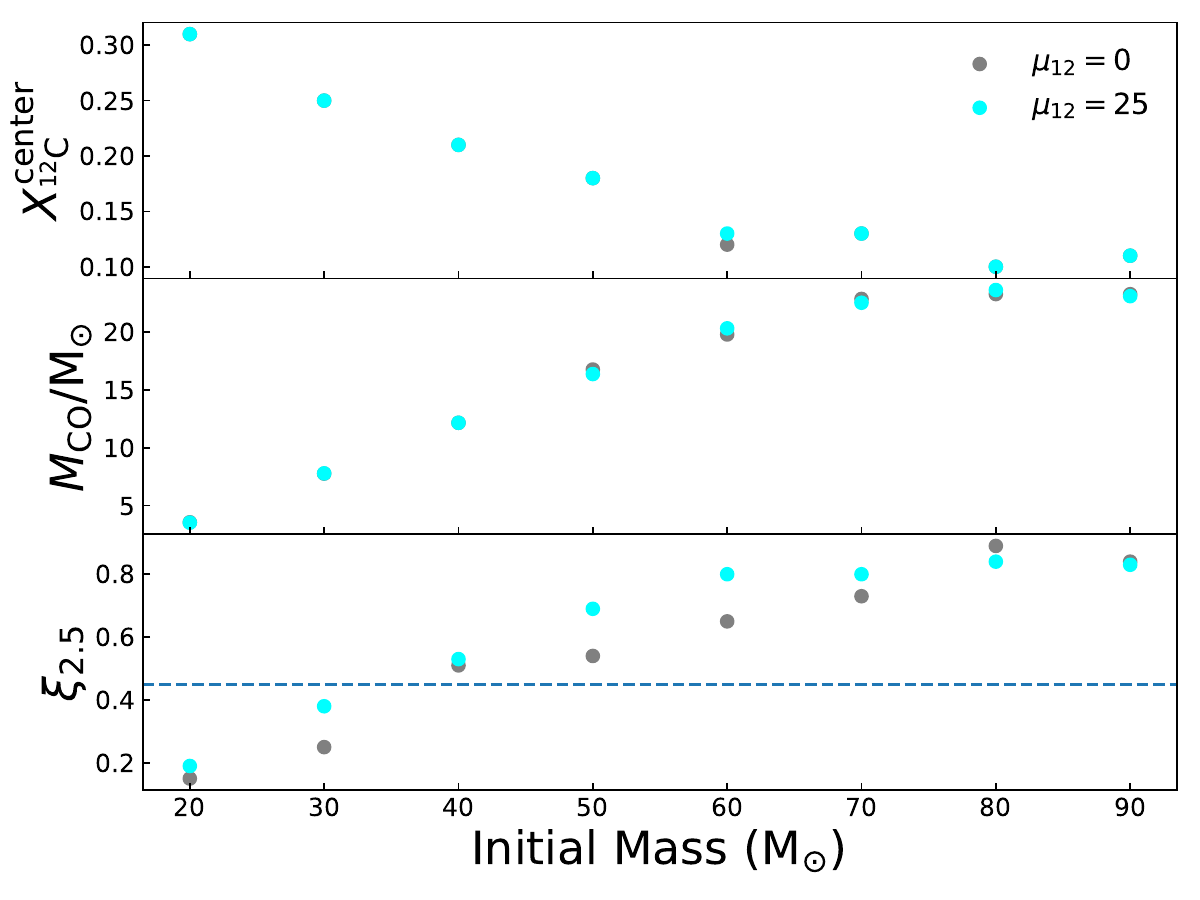}
    \vspace{-0.6cm}
    \caption{The change of $X_{^{12}\mathrm{C}}^{\mathrm{center}}$, $M_\mathrm{{CO}}/\mathrm{M_{\odot}}$, and $\xi_{2.5}$ when the $\rm \mu_{-12}=0$ and $\rm \mu_{-12}=25$, respectively. 
     $\rm \mu_{-12}$ is the NMM in units of $\rm 10^{-12} \mu_{B}$, FeCC is Fe Core Collapse, $X_{^{12}\mathrm{C}}^{\mathrm{center}}$ is the faction of central $^{12}\mathrm{C}$ at the end of central helium depletion, $M_\mathrm{{CO}}/\mathrm{M_{\odot}}$ is the CO core mass, $\xi_{2.5}$ is the compactness parameter. According to \citet{O'Connor2011}, the dashed line at $\xi_{2.5}=0.45$ serves as a demarcation, separating models that could potentially undergo explosion (below the line) or implode (above the line). The initial mass of stars ranges from 20 to 90 M$_{\odot}$ with a mass interval of 10 M$_\odot$.}    
    \label{fig:11}
\end{figure}

For further study, we discuss the effect of NMM on the compactness of presupernove stellar cores, which is usually expressed in $\xi_{\mathrm{M}}$. It is defined by \citep{O'Connor2011}

\begin{align}
    &&& \xi_{\mathrm{M}}=\frac{M / \mathrm{M}_{\odot}}{R\left(\mathrm{M}_{\text {bary }}=M\right) / 1000 \mathrm{~km}},
\end{align}
 
where $\rm M = 2.5\ M_{\odot}$ is used for quantifing the mean density near the iron core. In order to maintain substantial accuracy of compactness parameter, we end the evolution of massive stars when the collapse speed first reaches $\rm 1000\ km\ s^{-1}$ \citep{Sukhbold2014}. 

The extra cooling of massive stars can have an impact on the nuclear reactions inside the star, ultimately affecting the mass of the final CO core, as stated by \citep{Heger2008}. However, these changes are trivial for star with a mass larger than 20 M$_\odot$. Figure \ref{fig:11} shows the effects of NMM on $X(^{12}\mathrm{C})$ after He depletion, CO core mass, and $\xi_{2.5}$. It clearly explains that for stars with a mass larger than 20 M$_\odot$, even after considering additional cooling, the structure of massive stars does not undergo significant changes, so the explodability of the massive stars remains unchanged. 

\section{conclusions}
Using MESA, we simulate the stellar structure and evolution for massive stars with a mass from 20 to 90 M$_\odot$, and calculate the neutrino luminosity and give neutrino H-R diagram. Our results show: the greater the mass of the star, the more intense the nuclear reactions inside it are, and the higher the neutrino luminosity is. During helium burning, the neutrino luminosity decreases more significantly as the initial mass increases due to mass loss. During the hydrogen and helium burning stages, neutrinos are mainly produced by nuclear processes. Compared with photons, neutrinos carry away less energy. After carbon burning, neutrinos are mainly produced by thermal process (photo neutrino, pair annihilation and plasmon decay), meanwhile, photons are trapped, neutrinos dominate stellar energy loss. However, after Si burning, the large number of weak processes allows nuclear reaction processes to produce more neutrinos than thermal processes.

Low metallicity can facilitate a more condensed structure within the massive star, leading to higher $T_{\mathrm{c}}$ and higher neutrino luminosities during hydrogen burning. During helium burning, low mass loss due to low metallicity can prevent a significant decrease in neutrino luminosity due to neutrino release from the burning of a large number of hydrogen shells.

Rotation also affects the neutrino luminosity, which is determined by the initial mass and metallicity. At the high metallicity ($Z=0.014$), for stars with a mass lower than 40 M$_\odot$, rotational mixing can increase neutrino luminosity during hydrogen, C, and O burning. For stars with a mass higher than 40 M$_\odot$, the rotational mixing is less efficient due to mass loss, neutrino luminosity does not increase much. For more massive stars, neutrino luminosity even decreases during C and O burning, such as our 90 M$_{\odot}$ star model. At low metallicity, where mass loss is negligible, the neutrino luminosity of all massive stars with high rotation can increase significantly during hydrogen, C, and O burning. 

The energy loss caused by the NMM does not have effects on the evolutionary destiny of stars with a mass larger than 20 M$_\odot$, and does not significant change the compactness at FeCC. Therefore, the cooling via NMM has no effect on the explodability of massive stars. 

\section*{Acknowledgements}
This work received the generous support of 
the National Natural Science Foundation of 
China, project No.U2031204, 12163005, 12288102, and 12263006, the 
science research grants from the China Manned
Space Project with NO.CMS-CSST-2021-A10, the
Natural Science Foundation of Xinjiang No.
2021D01C075, 2022D01D85, and 2022TSYCLJ0006.

\section*{Data Availability}
The data generated, analysed and presented in this study are available from the 
corresponding authors on reasonable request.



\bibliographystyle{mnras}
\bibliography{library1} 





\bsp	
\label{lastpage}
\end{document}